\documentstyle[epsfig,12pt]{article}

\begin{document}


\mbox{ } \\[-1cm]
\mbox{ }\hfill DESY 00--001\\
\mbox{ }\hfill IFT--00/03\\
\mbox{ }\hfill KIAS--P00004\\
\mbox{ }\hfill PM/00--02\\
\mbox{ }\hfill SNUTP 00--003\\
\mbox{ }\hfill hep--ph/0002033\\
\mbox{ }\hfill \today\\

\begin{center}
{\Large{\bf RECONSTRUCTING THE CHARGINO SYSTEM \\[2mm]
            AT \boldmath{$e^+e^-$} LINEAR COLLIDERS}}\\[1cm]
S.~Y.~Choi$^1$, A.~Djouadi$^2$, M.~Guchait$^3$, J.~Kalinowski$^4$, 
H.S.~Song$^5$ and P.~M.~Zerwas$^3$ 
\end{center}

\bigskip

{\small
\begin{enumerate}
\item[{}] $^1$ Korea Institute for Advanced Study, Seoul 130--012, Korea
\item[{}] $^2$ Phys. Math. et Th\'{e}orique, Universit\'{e} Montpellier II,
               F--34095 Montpellier, France
\item[{}] $^3$ Deutsches Elektronen-Synchrotron DESY, D-22603 Hamburg, Germany
\item[{}] $^4$ Inst. Theor. Physics, Warsaw University, PL--00681 Warsaw, 
               Poland
\item[{}] $^5$ Center for Theor. Physics and Dep. of Physics, 
               Seoul National Univ., Seoul 151--742, Korea
\end{enumerate}
}
\bigskip

\begin{abstract}
In most supersymmetric theories charginos, $\tilde{\chi}^\pm_{1,2}$, belong 
to the class of the lightest supersymmetric particles.
The chargino system can be reconstructed completely in $e^+e^-$ collider 
experiments: $e^+e^-\rightarrow\tilde{\chi}_i^+ \tilde{\chi}_j^- \, \,  
[i,j=1,2]$. By measuring the total cross sections and the 
asymmetries with polarized beams, the chargino masses and the 
gaugino--higgsino mixing angles of these states can be determined  
accurately.  If only the lightest charginos $\tilde{\chi}_1^\pm$
are kinematically accessible in a first phase of the machine, transverse 
beam polarization  or the measurement of chargino polarization
in the final state is needed to determine the mixing angles. 
From these observables the fundamental SUSY parameters can be derived: 
the SU(2) gaugino mass $M_2$, 
the modulus and the cosine of the CP--violating phase of the higgsino mass 
parameter $\mu$, and $\tan\beta = v_2/v_1$, the ratio of 
the vacuum expectation values of the two neutral Higgs doublet fields. 
The remaining two--fold ambiguity of the phase can be resolved by
measuring the normal polarization of the charginos. Sum rules of the cross
sections can be exploited to investigate the closure of the two--chargino
system.
\end{abstract}
%



\renewcommand{\thefootnote}{\alph{footnote}}
\newpage

\section{INTRODUCTION}

In supersymmetric theories, the spin-1/2 partners of the $W^{\pm}$ gauge 
bosons and the charged Higgs bosons, $\tilde W^\pm$ and $\tilde H^\pm$,
mix to form chargino mass eigenstates $\chi_{1,2}^\pm$.
The chargino mass matrix \cite{R1} in the 
$(\tilde{W}^-,\tilde{H}^-)$ basis 
\begin{eqnarray}
{\cal M}_C=\left(\begin{array}{cc}
                M_2                &      \sqrt{2}m_W\cos\beta  \\
             \sqrt{2}m_W\sin\beta  &             \mu   
                  \end{array}\right)\
\label{eq:mass matrix}
\end{eqnarray}
is built up by the fundamental supersymmetry (SUSY) 
parameters: the SU(2) gaugino mass $M_2$, the higgsino mass parameter $\mu$, 
and the ratio $\tan\beta=v_2/v_1$ of the vacuum expectation values of the two 
neutral Higgs fields which break the electroweak symmetry. 
In CP--noninvariant theories, the mass parameters are complex \cite{R1}. 
However, by reparametrization of the fields, 
$M_2$ can be assumed real and positive without loss of generality so that
the only non--trivial reparametrization--invariant phase 
may be attributed to $\mu$:
\begin{eqnarray}
\mu=|\mu|\,{\rm e}^{i\Phi_\mu}\ \ {\rm with}\ \ 0 \leq \Phi_\mu \leq 2\pi
\end{eqnarray}
Once charginos will have been discovered, the experimental analysis of their 
properties in production and decay mechanisms will reveal the basic structure
of the underlying supersymmetric theory.\\

Charginos are produced in $e^+e^-$ collisions, either in diagonal or in
mixed pairs \cite{R2}-\cite{10A}: 
\begin{eqnarray*}
e^+ e^- \ \rightarrow \ \tilde{\chi}^+_i \ \tilde{\chi}^-_j 
        \ \ \ [\,i,j=1,2\,]
\,
\end{eqnarray*}

\noindent
Depending on the collider energy and the chargino masses, the
following scenarios will be analyzed:\\

\noindent
{\bf (i)} If the energy in the first phase of the machine is only
sufficient to produce the light chargino pair 
$\tilde{\chi}^+_1\tilde{\chi}^-_1$, the underlying fundamental parameters, 
up to at most two--fold ambiguity, 
can be extracted from the mass $m_{\tilde{\chi}^\pm_1}$, the total
production cross section and the measurement of longitudinal left--right
and transverse asymmetries. 
Alternatively to beam polarization, the polarization of the charginos
in the final state may be exploited. The $\tilde{\chi}^\pm$ polarization 
vectors 
and $\tilde{\chi}^+$--$\tilde{\chi}^-$ spin--spin correlation tensor can be
determined from the decay distributions of the charginos. 
We will assume that the charginos decay into the lightest neutralino
$\tilde{\chi}^0_1$, which is taken to be stable, and a pair of quarks
and antiquarks or leptons:
$\tilde{\chi}^\pm_1\rightarrow \tilde{\chi}^0_1\,f\bar{f}'$. 
No detailed information on the decay
dynamics, nor on the structure of the neutralino, is needed to carry
out the spin analysis \cite{R4A}.\\

\noindent
{\bf (ii)} If the collider energy is sufficient to produce the two chargino
states in pairs, the underlying fundamental SUSY parameters $\{M_2,|\mu|, 
\cos\Phi_{\mu}, 
\tan\beta\}$ can be extracted {\it unambiguously} from the masses
$m_{\tilde{\chi}^\pm_{1,2}}$, the total production cross sections, and the
left--right (LR) asymmetries with polarized electron beams, while the
phase $\Phi_\mu$ is determined up to a two--fold ambiguity
$\Phi_\mu\leftrightarrow 2\pi-\Phi_\mu$. As shown 
in Ref.\cite{R11}, this ambiguity can be resolved
by measuring manifestly CP--noninvariant observables 
associated with the normal polarization of the charginos.\\

These analyses of the chargino sector are independent of the structure of the
neutralino sector \cite{KM}. While the structure of the chargino sector 
in large classes of supersymmetric theories is isomorphic to the minimal 
supersymmetric standard model (MSSM), we expect the neutralino
sector to be more complex in general, reflecting the complexity of a Higgs
sector extended beyond the minimal form.\\

The analysis will be based strictly on low--energy SUSY. 
To clarify the analytical structure, the reconstruction of the basic SUSY 
parameters presented here
is carried out at the tree level; the small loop  corrections \cite{DKR} 
include 
parameters from other sectors of the MSSM demanding iterative higher--order
expansions in global analyses at the very end. Once
these basic parameters will have been extracted experimentally, they may be
confronted, for instance, with the ensemble of  relations predicted in 
Grand Unified Theories.\\

In this report we present a coherent and comprehensive
description of the chargino system at $e^+e^-$ linear colliders,
based on scattered elements discussed earlier in Refs.~\cite{A1}--\cite{A3}.
The report will be divided into six parts. In
Section~2 we recapitulate the central elements of 
the mixing formalism for the charged gauginos and higgsinos. 
In Section~3 the cross sections for chargino production, the left--right 
asymmetries, and the polarization vectors of the charginos are given. 
In Section~4 we describe a phenomenological analysis of the
light $\tilde{\chi}^\pm_1$ states based on a specific scenario to exemplify 
the procedure for extracting the fundamental SUSY parameters in a 
model--independent way.  In Section~5 the analysis is extended to the
complete set $\tilde{\chi}^\pm_{1,2}$ of chargino states, leading to
an unambiguous determination of the SU(2) gaugino parameters. Conclusions
are given in Section~6.

\section{MIXING FORMALISM}
\label{sec:mixing}

In the MSSM and many of its
extensions, the two charginos $\tilde{\chi}^\pm_{1,2}$ are mixtures
of the charged SU(2) gauginos and higgsinos.
As a consequence of possible field redefinitions, the parameters
$\tan\beta$ and $M_2$ can be chosen  
real and positive. 
Since the chargino mass matrix ${\cal M}_C$ is not symmetric, two
different 
unitary matrices acting on the left-- and right--chiral 
$(\tilde{W},\tilde{H})_{L,R}$ two--component 
states 
\begin{eqnarray}
U_{L,R}\left(\begin{array}{c}
             \tilde{W}^- \\
             \tilde{H}^-
             \end{array}\right)_{L,R} =
       \left(\begin{array}{c}
             \tilde{\chi}^-_1 \\
             \tilde{\chi}^-_2
             \end{array}\right)_{L,R} 
\end{eqnarray}
are needed to diagonalize the matrix eq.(\ref{eq:mass matrix}).
The unitary matrices $U_L$ and $U_R$ can be parameterized in the
following way \cite{R11}:
\begin{eqnarray}
&& U_L=\left(\begin{array}{cc}
             \cos\phi_L & {\rm e}^{-i\beta_L}\sin\phi_L \\
            -{\rm e}^{i\beta_L}\sin\phi_L & \cos\phi_L
             \end{array}\right) \nonumber\\
&& U_R=\left(\begin{array}{cc}
             {\rm e}^{i\gamma_1} & 0 \\
             0 & {\rm e}^{i\gamma_2}
             \end{array}\right)
        \left(\begin{array}{cc}
             \cos\phi_R & {\rm e}^{-i\beta_R}\sin\phi_R \\
            -{\rm e}^{i\beta_R}\sin\phi_R & \cos\phi_R
             \end{array}\right) 
\end{eqnarray}
The mass eigenvalues $m^2_{\tilde{\chi}^\pm_{1,2}}$ are given by
\begin{eqnarray}
m^2_{\tilde{\chi}^\pm_{1,2}}
   =\frac{1}{2}\left[M^2_2+|\mu|^2+2m^2_W\mp \Delta_C\right]
\end{eqnarray}
with $\Delta_C$ involving the phase  $\Phi_\mu$: 
\begin{eqnarray}
\Delta_C=\sqrt{(M^2_2-|\mu|^2)^2+4m^4_W\cos^2 2\beta
              +4m^2_W(M^2_2+|\mu|^2)+8m^2_WM_2|\mu|
               \sin2\beta\cos\Phi_\mu}
\end{eqnarray}
The quantity $\Delta_C$ determines the difference of the two chargino 
masses: $\Delta_C = m^2_{\tilde{\chi}^\pm_2} - m^2_{\tilde{\chi}^\pm_1}$.
The four phase angles $\{\beta_L,\beta_R,\gamma_1,\gamma_2\}$
are not independent but can be expressed in terms of  the invariant angle
$\Phi_\mu$:
\begin{eqnarray}
&& \tan\beta_L=-\frac{\sin\Phi_\mu}{\cos\Phi_\mu
                                      +\frac{M_2}{|\mu|}\cot\beta}
   \qquad \hskip 0.95cm
   \tan\beta_R=+\frac{\sin\Phi_\mu}{\cos\Phi_\mu
                                      +\frac{M_2}{|\mu|}\tan\beta}
   \nonumber\\
&& \tan\gamma_1=+\frac{\sin\Phi_\mu}{\cos\Phi_\mu
                   +\frac{M_2\left[m^2(\tilde{\chi}^\pm_1)-|\mu|^2\right]}{
                          |\mu| m^2_W\sin 2\beta}} 
   \qquad
   \tan\gamma_2=-\frac{\sin\Phi_\mu}{\cos\Phi_\mu
                   +\frac{M_2m^2_W\sin 2\beta}{
                         |\mu|\left[m^2(\tilde{\chi}^\pm_2)-M^2_2\right]}}
\label{eq:four phases}
\end{eqnarray}
All four phase angles vanish in CP--invariant theories for which 
$\Phi_\mu = 0$ or $\pi$.  The rotation angles 
$\phi_L$ and $\phi_R$ satisfy the relations:
\begin{eqnarray}
&&\cos 2\phi_{L,R}=-\left[M_2^2-|\mu|^2\mp 2m^2_W\cos 2\beta\right]/\Delta_C
   \nonumber\\ 
&&\sin 2\phi_{L,R}=-2m_W\sqrt{M^2_2+|\mu|^2\pm(M^2_2-|\mu|^2)\cos 2\beta
                     +2M_2|\mu|\sin2\beta\cos\Phi_\mu}/\Delta_C
\end{eqnarray}

The two rotation angles $\phi_{L,R}$ and the phase angles
$\{\beta_L,\beta_R,\gamma_1,\gamma_2\}$ define the couplings
of the chargino--chargino--$Z$ vertices:
{\small
\begin{eqnarray}
&&{ }\hskip -5mm  \langle\tilde{\chi}^-_{1L}|Z|\tilde{\chi}^-_{1L}\rangle 
  = -\frac{g_W}{c_W} \left[s_W^2 - \frac{3}{4}-\frac{1}{4}
     \cos 2\phi_L\right] \,\qquad 
   \langle\tilde{\chi}^-_{1R}|Z|\tilde{\chi}^-_{1R}\rangle 
  = -\frac{g_W}{c_W} \left[s_W^2-\frac{3}{4}-\frac{1}{4}\cos 
    2\phi_R\right] \nonumber\\
&&{ }\hskip -5mm  \langle\tilde{\chi}^-_{1L}|Z|\tilde{\chi}^-_{2L}\rangle 
  = +\frac{g_W}{4c_W}\,{\rm e}^{-i\beta_L}\sin 2\phi_L\qquad 
  \qquad\quad\quad  \langle\tilde{\chi}^-_{1R}|Z|\tilde{\chi}^-_{2R}\rangle 
  = +\frac{g_W}{4c_W}\,{\rm e}^{-i(\beta_R-\gamma_1+\gamma_2)}
                        \sin 2\phi_R\nonumber\\
&&{ }\hskip -5mm  \langle\tilde{\chi}^-_{2L}|Z|\tilde{\chi}^-_{2L}\rangle 
  = -\frac{g_W}{c_W} \left[s_W^2 - \frac{3}{4}+\frac{1}{4}
     \cos 2\phi_L\right] \,\qquad
\langle\tilde{\chi}^-_{2R}|Z|\tilde{\chi}^-_{2R}\rangle 
 = -\frac{g_W}{c_W} \left[s_W^2-\frac{3}{4}+\frac{1}{4}\cos 
    2\phi_R\right]\nonumber 
\end{eqnarray}
}
and the electron--sneutrino--chargino vertices:
\begin{eqnarray}
&&{ }\hskip -5mm  \langle\tilde{\chi}^-_{1R}|\tilde{\nu}|e^-_L\rangle 
  = -g_{\tilde{W}}\,{\rm e}^{i\gamma_1}\cos\phi_R \nonumber\\[1mm]
&&{ }\hskip -5mm  \langle\tilde{\chi}^-_{2R}|\tilde{\nu}|e^-_L\rangle 
  = +g_{\tilde{W}}\,{\rm e}^{i(\beta_R+\gamma_2)}\sin\phi_R
\label{eq:vertex}
\end{eqnarray}
with $s_W^2 =1-c_W^2 \equiv \sin^2\theta_W$ denoting the electroweak mixing
angle. $g_W$ and $g_{\tilde{W}}$ are the $e\nu W$ gauge coupling and the 
$e\tilde{\nu} \tilde{W}$ Yukawa coupling, respectively. 
They are identical in supersymmetric theories:
\begin{eqnarray}
g_{\tilde{W}}=g_W=e/s_W
\end{eqnarray}
Since the coupling to the higgsino component, which is proportional
to the electron mass, can be neglected in the sneutrino vertex,
the sneutrino couples only to left--handed electrons. 
The diagonal and L/R symmetric photon--chargino vertices are as usual
\begin{eqnarray}
\langle\tilde{\chi}^-_i|\gamma|\tilde{\chi}^-_i\rangle = e
\end{eqnarray}
CP--violating effects
are manifest only in mixed $\tilde{\chi}_1\tilde{\chi}_2$ pairs.\\

Conversely, the fundamental SUSY parameters $M_2$, $|\mu|$, $\tan\beta$ 
and the phase parameter $\cos\Phi_\mu$ can be extracted from the
chargino $\tilde{\chi}^\pm_{1,2}$ parameters: the masses 
$m_{\tilde{\chi}^\pm_{1,2}}$ and the two mixing angles $\phi_L$ and 
$\phi_R$ of the left-- and right--chiral components of the wave function 
(see Sect.~\ref{fundamental}).\\ 

\section{CHARGINO PRODUCTION IN $e^+e^-$ COLLISIONS}

The production of chargino pairs at $e^+e^-$ colliders is
based on three mechanisms: $s$--channel $\gamma$ 
and $Z$ exchanges, and $t$--channel $\tilde{\nu}_e$ exchange, cf. Fig.1. 
The transition matrix element, after a Fierz transformation of the 
$\tilde{\nu}_e$--exchange amplitude,
\begin{eqnarray}
T[e^+e^-\rightarrow\tilde{\chi}^-_i\tilde{\chi}^+_j]
  =\frac{e^2}{s}Q_{\alpha\beta}
   \left[\bar{v}(e^+)\gamma_\mu P_\alpha  u(e^-)\right]
   \left[\bar{u}(\tilde{\chi}^-_i) \gamma^\mu P_\beta 
               v(\tilde{\chi}^+_j) \right]
\label{eq:production amplitude}
\end{eqnarray}
can be expressed in terms of four bilinear charges, defined by  
the chiralities $\alpha,\beta=L,R$ of the associated 
lepton and chargino currents. After introducing the following 
notation,
\begin{eqnarray}
&& \hskip -1.9cm 
   D_L=1+\frac{D_Z}{s_W^2 c_W^2}(s_W^2 -\frac{1}{2})(s_W^2-\frac{3}{4})\,
       \quad\qquad 
   F_L=\frac{D_Z}{4s_W^2 c_W^2}(s^2_W-\frac{1}{2})\ \nonumber\\
&& \hskip -1.9cm 
   D_R=1+\frac{D_Z}{c_W^2}(s_W^2-\frac{3}{4})\, \quad \quad \quad 
   \quad\qquad\qquad 
   F_R=\frac{D_Z}{4c_W^2}
\label{eq:DFLR}
\end{eqnarray}
and
\begin{eqnarray}
D'_L=D_L+\left(\frac{g_{\tilde{W}}}{g_W}\right)^2\, 
              \frac{D_{\tilde{\nu}}}{4s^2_W}
   \qquad\qquad\hskip 1.3cm 
F'_L=\,F_L-\,\left(\frac{g_{\tilde{W}}}{g_W}\right)^2\, 
              \frac{D_{\tilde{\nu}}}{4s^2_W}
\end{eqnarray}
the four bilinear charges $Q_{\alpha\beta}$ are linear in the 
mixing parameters $\cos2\phi_{L,R}$ and $\sin2\phi_{L,R}$; for the diagonal 
$\tilde{\chi}^-_1\tilde{\chi}^+_1$, $\tilde{\chi}^-_2\tilde{\chi}^+_2$
modes and the mixed mode $\tilde{\chi}^-_1\tilde{\chi}^+_2$ we find:\\
\begin{eqnarray}
\{11\}/\{22\} : &&{ } \hskip -2mm Q_{LL}=D_L\mp F_L\cos 2\phi_L\, 
                 \hskip 1.01cm \quad \qquad
                 Q_{RL}=D_R\mp F_R\cos 2\phi_L\, \nonumber \\
            &&{ } \hskip -2mm Q_{LR}=D'_L\mp F'_L \cos 2\phi_R\, 
                    \quad \qquad\hskip 0.96cm 
                 Q_{RR}=D_R\mp F_R\cos 2\phi_R \\
            &&                                \nonumber\\
{ } 
\{12\}/\{21\}:  
            &&{ }\hskip -2mm Q_{LL}=F_L\,{\rm e}^{\mp i\beta_L}\sin 2\phi_L\, 
                 \quad \quad \quad \quad\,\hskip 7mm
                 Q_{RL}=F_R\,{\rm e}^{\mp i\beta_L}\sin 2\phi_L\, \nonumber\\ 
            && { }\hskip  -2mm Q_{LR}=F'_L\,{\rm e}^{\mp i(\beta_R-\gamma_1
                +\gamma_2)}\sin 2\phi_R\, \quad\hskip 5.6mm
                 Q_{RR}=F_R\,{\rm e}^{\mp i(\beta_R-\gamma_1
                 +\gamma_2)}\sin 2\phi_R\,
\label{eq:[12]}
\end{eqnarray}
The first index in $Q_{\alpha \beta}$
refers to the chirality of the $e^\pm$ current, the second index to the 
chirality of the $\tilde{\chi}^\pm$ current. The $\tilde{\nu}$ 
exchange affects only the $LR$ chirality charge $Q_{LR}$ 
while all other amplitudes 
are built up by $\gamma$ and/or $Z$ exchanges only. The first term in 
$D_{L,R}$ is generated by the $\gamma$ exchange;
$D_Z=s/(s-m^2_Z+im_Z\Gamma_Z)$ denotes the $Z$ propagator and
$D_{\tilde{\nu}} = s/(t- m_{\tilde{\nu}}^2)$ the $\tilde{\nu}$ propagator
with momentum transfer $t$. The non--zero $Z$ width can in general be 
neglected for the energies considered 
in the present analysis so that the charges are rendered complex in 
the Born approximation only through the 
CP--noninvariant phase.\\   

For the sake of convenience we introduce eight quartic charges for each of 
the production processes of the diagonal and mixed chargino pairs,
respectively. These charges \cite{R7} correspond to independent helicity 
amplitudes
which describe the chargino production processes for polarized
electrons/positrons with negligible lepton masses. Expressed in terms of 
bilinear charges they are collected in Table 1, including the
transformation properties under P and CP.\\
\begin{table*}[\hbt]
\caption[{\bf Table 1:}]{\label{tab:quartic} 
{\it The independent quartic charges of the chargino system, the measurement 
     of which determines the chargino mass matrix.}}
\begin{center}
\begin{tabular}{|c|c|l|}\hline
 &  &  \\[-4mm]
${\rm P}$ & ${\rm CP}$ & { }\hskip 2cm Quartic charges \\\hline \hline
 &  &  \\[-3mm]
 even    &  even     & $Q_1 =\frac{1}{4}\left[|Q_{RR}|^2+|Q_{LL}|^2
                       +|Q_{RL}|^2+|Q_{LR}|^2\right]$ \\[2mm]
         &           & $Q_2 = \frac{1}{2}{\rm Re}\left[Q_{RR}Q^*_{RL}
                       +Q_{LL}Q^*_{LR}\right]$ \\[2mm]
         &           & $Q_3 = \frac{1}{4}\left[|Q_{RR}|^2+|Q_{LL}|^2
                       -|Q_{RL}|^2-|Q_{LR}|^2\right]$ \\[2mm]
         &           & $Q_5=\frac{1}{2}{\rm Re} \left[Q_{LR}Q^*_{RR}
                       +Q_{LL}Q^*_{RL}\right]$ \\
 & & \\[-3mm]
\cline{2-3} 
 & & \\[-3mm]
         &  odd      & $Q_4=\frac{1}{2}{\rm Im}\left[Q_{RR}Q^*_{RL}
                       +Q_{LL}Q^*_{LR}\right]$\\[2mm] \hline \hline
 & & \\[-3mm]
 odd     &  even     & $Q'_1=\frac{1}{4}\left[|Q_{RR}|^2+|Q_{RL}|^2
                        -|Q_{LR}|^2-|Q_{LL}|^2\right]$\\[2mm]
         &           & $Q'_2=\frac{1}{2}{\rm Re}\left[Q_{RR}Q^*_{RL}
                        -Q_{LL}Q^*_{LR}\right]$ \\[2mm]
         &           & $Q'_3=\frac{1}{4}\left[|Q_{RR}|^2+|Q_{LR}|^2
                        -|Q_{RL}|^2-|Q_{LL}|^2\right]$\\[2mm] 
\hline
\end{tabular}
\end{center}
\end{table*}

The charges $Q_1$ to $Q_5$ are manifestly parity--even,
$Q'_1$ to $Q'_3$ are parity--odd. The charges 
$Q_1$ to $Q_3$, $Q_5$, and $Q'_1$ to $Q'_3$ are CP--invariant\footnote{ 
When expressed in terms of the fundamental SUSY parameters, 
these charges do depend nevertheless on $\cos\Phi_\mu$ indirectly through 
$\cos 2\phi_{L,R}$, in the same way as the $\tilde{\chi}^\pm_{1,2}$
masses depend indirectly on 
this parameter.}.
$Q_4$ changes sign under CP transformations\footnote{The 
P--odd and CP--even/CP--odd counterparts to $Q_5/Q_4$, which carry a 
negative sign between the corresponding $L$ and $R$ components, do not 
affect the observables under consideration.}, yet depends only on one
combination $(\beta_L-\beta_R+\gamma_1-\gamma_2)$ of the CP angles.
The CP invariance of $Q_2$ and $Q'_2$ can easily be proved by noting that
\begin{eqnarray}
2 m_{\tilde\chi_1^\pm} m_{\tilde\chi_2^\pm} 
  \cos(\beta_L-\beta_R+\gamma_1-\gamma_2)
  \sin2\phi_L\sin2\phi_R \nonumber\\    
= (m^2_{\tilde\chi^\pm_1}+m^2_{\tilde\chi^\pm_2})
      \left(1-\cos2\phi_L \cos2\phi_R\right)-4m^2_W 
\label{eq:sisisi} 
\end{eqnarray}
Therefore, all the production cross sections 
$\sigma[e^+e^-\rightarrow \tilde{\chi}^+_i\tilde{\chi}^-_j]$ for any 
combination of pairs $\tilde{\chi}^+_i\tilde{\chi}^-_j$ depend only on  
$\cos 2 \phi_L$
and $\cos 2 \phi_R$ apart from the chargino masses, the sneutrino
mass and the Yukawa couplings. For longitudinally--polarized electron beams, 
the sums and differences of the quartic charges are restricted to either 
$L$ or $R$ components (first index) of the $e^\pm$ currents.\\ 

Defining the $\tilde{\chi}^-_i$ production 
angle with respect to the electron flight--direction by the polar angle 
$\Theta$ and the azimuthal angle $\Phi$ with respect to the electron 
transverse polarization, the helicity amplitudes 
can be derived from eq.(\ref{eq:production amplitude}). While electron
and positron helicities are opposite to each other in all amplitudes,
the $\tilde{\chi}^-_i$ and $\tilde{\chi}^+_j$ helicities are in
general not correlated due to the non--zero masses of the particles;
amplitudes with equal $\tilde{\chi}^-_i$ and $\tilde{\chi}^+_j$
helicities are reduced only to order  
$\propto m_{\tilde{\chi}^\pm_{i,j}} /\sqrt{s}$ 
for asymptotic energies. The helicity amplitudes
may be expressed as  
$T_{ij}(\sigma;\lambda_i,\lambda_j)=2\pi\alpha\,{\rm e}^{i\sigma\Phi}\langle
\sigma;\lambda_i\lambda_j\rangle$, 
denoting the electron helicity by the first index $\sigma$, the
$\tilde{\chi}^-_i$ and $\tilde{\chi}^+_j$ helicities by the remaining
two indices, $\lambda_i$ and $\lambda_j$, respectively.
The explicit form of the helicity amplitudes  
$\langle\sigma;\lambda_i\lambda_j\rangle$ can be found in 
Ref.~\cite{A2}.\\

\subsection{Production cross sections}

Since the gaugino and higgsino interactions depend on the chirality of
the states, the polarized electron and positron beams are powerful tools 
to reveal the composition of charginos.  
To describe the electron and positron polarizations, 
the reference frame must be fixed. The electron--momentum direction
will define the $z$--axis and the electron transverse polarization--vector 
the $x$--axis.
The azimuthal angle of the transverse polarization--vector of the positron
is called $\eta$ with respect to the $x$--axis. In this notation, 
the polarized differential cross section 
is given in terms of the electron and positron polarization vectors 
$P$=$(P_T,0,P_L)$ and $\bar{P}$=$(\bar{P}_T \cos\eta,\bar{P}_T\sin\eta,
-\bar{P}_L)$ by
\begin{eqnarray}
\frac{{\rm d}\sigma}{{\rm d}\Omega}
  =\frac{\alpha^2}{16 s} \lambda^{1/2} \bigg[
     (1-P_L\bar{P}_L)\,\Sigma_{\rm unp}+(P_L-\bar{P}_L)\,\Sigma_{LL}
  +P_T\bar{P}_T\cos(2\Phi-\eta)\,\Sigma_{TT}\bigg]\
\end{eqnarray}
with the coefficients $\Sigma_{\rm unp}$, $\Sigma_{LL}$, $\Sigma_{TT}$
depending 
only on the polar angle $\Theta$, but not on the azimuthal 
angle $\Phi$ any more;
$\lambda=[1-(\mu_i+\mu_j)^2][1-(\mu_i-\mu_j)^2]$ is the two--body 
phase space function, and $\mu_i^2=m^2_{\tilde\chi_i^\pm}/s$.
The coefficients $\Sigma_{\rm unp}$, $\Sigma_{LL}$, and $\Sigma_{TT}$
can be expressed in terms of the quartic charges:
\begin{eqnarray}
\Sigma_{\rm unp}&=& 4\bigg\{\left[1-(\mu^2_i - \mu^2_j)^2
                   +\lambda\cos^2\Theta\right]Q_1
                   +4\mu_i\mu_j Q_2+2\lambda^{1/2} Q_3\cos\Theta\bigg\}
                  \nonumber\\
\Sigma_{LL}     &=& 4\bigg\{\left[1-(\mu^2_i - \mu^2_j)^2
                   +\lambda\cos^2\Theta\right]Q'_1
                   +4\mu_i\mu_j Q'_2+2\lambda^{1/2} Q'_3\cos\Theta\bigg\}
                  \nonumber\\
\Sigma_{TT}     &=&-4\lambda \sin^2\Theta\,\, Q_5
\label{eq:initial}
\end{eqnarray}
If the production angles could be measured unambiguously on an event--by--event
basis, the quartic charges could be extracted directly from the angular 
dependence of the cross section at a single energy. However, since charginos 
decay into the invisible lightest neutralinos and SM fermion pairs, 
the production angles cannot be determined completely
on an event--by--event basis.
The transverse distribution can be extracted by using an
appropriate weight function for the azimuthal angle $\Phi$. 
This leads us to the 
following integrated polarization--dependent cross sections
as physical observables:
\begin{eqnarray}
\sigma_R&=&\int{\rm d}\Omega\,\,\frac{{\rm d}\sigma}{{\rm d}\Omega}
              \left[P_L=-\bar{P}_L=+1\right] \nonumber\\
\sigma_L&=&\int{\rm d}\Omega\,\,\frac{{\rm d}\sigma}{{\rm d}\Omega}
              \left[P_L=-\bar{P}_L=-1\right] \nonumber\\
\sigma_T&=&\int{\rm d}\Omega\,\,\left(\frac{\cos 2\Phi}{\pi}\right)
               \frac{{\rm d}\sigma}{{\rm d}\Omega}
              \left[P_T=\bar{P}_T=1;\,\eta=\pi\right]
\label{eq:xsections}
\end{eqnarray}
As a result, nine independent physical observables can be constructed at
a given c.m. energy by
means of beam polarization in the three production processes; three in each 
mode $\{ij\}=\{11\},\{12\}$ and \{22\}.\\

\subsection{Chargino polarization and spin  correlations}

If the lepton beams are not polarized, the chiral structure
of the charginos can be inferred from the polarization of the 
$\tilde{\chi}^-_i\tilde{\chi}^+_j$ pairs produced in $e^+e^-$ 
annihilation.\\

The polarization vector $\vec{\cal P}=({\cal P}_T,{\cal P}_N,
{\cal P}_L)$ is defined in the rest frame of the particle $\tilde{\chi}^-_i$. 
${\cal P}_L$ denotes the component parallel to the $\tilde{\chi}^-_i$ flight 
direction in the c.m. frame, ${\cal P}_T$ the transverse component in the 
production plane, and ${\cal P}_N$ the  component normal to the production 
plane. 
The longitudinal and transverse components of the $\tilde{\chi}_i^-$
polarization vector can easily be expressed in terms of the
quartic charges:
%
{\small
\begin{eqnarray}
&& {\cal P}_L =4\left\{ 2(1-\mu^2_i-\mu^2_j)\,\cos\Theta\,{Q'}_1
              +4\mu_i\mu_j\,\cos\Theta\, {Q'}_2
              +\lambda^{1/2}[1+\cos^2\Theta-(\mu^2_i-\mu^2_j)]\, {Q'}_3 
              \right\} / {\cal N} \nonumber \\
&& {\cal P}_T =-8\sin\Theta\left\{[(1-\mu^2_i+\mu^2_j)\,{Q'}_1
              +\lambda^{1/2}\, {Q'}_3\cos\Theta]\mu_i
              +(1+\mu^2_i-\mu^2_j)\mu_j\, {Q'}_2\right\}/{\cal N}\nonumber \\ 
&& {\cal P}_N =8\lambda^{1/2}\mu_j\,\sin\Theta\, Q_4/{\cal N}
\end{eqnarray}
}
with the normalization ${\cal N}$ given by  
\begin{eqnarray}
{\cal N} = 4\left\{\left[1-(\mu^2_i - \mu^2_j)^2 +\lambda\cos^2\Theta\right]Q_1
          +4\mu_i\mu_j Q_2+2\lambda^{1/2} Q_3\cos\Theta\right\}
\end{eqnarray}

The normal component ${\cal P}_N$ can only be generated by complex production
amplitudes.  Non-zero phases are present in the fundamental
supersymmetric parameters if CP is broken in the
supersymmetric interaction \cite{R1}. Also, the non--zero width of the
$Z$ boson and loop corrections generate non--trivial phases; however,
the width effect is negligible for high energies and the effects due
to radiative corrections are small. Neglecting loops and the small
$Z$--width, the normal $\tilde{\chi}^-_1$ and $\tilde{\chi}^+_1$
polarizations in $e^+e^-\rightarrow \tilde{\chi}^-_1 \tilde{\chi}^+_1$
are zero since the $\tilde{\chi}_1 \tilde{\chi}_1 \gamma$ and
$\tilde{\chi}_1 \tilde{\chi}_1 Z$ vertices are real even for non-zero
phases in the chargino mass matrix, and the sneutrino--exchange
amplitude is real, too. The same holds true for $\tilde{\chi}^-_2
\tilde{\chi}^+_2$ production. Only for nondiagonal
$\tilde{\chi}^-_1 \tilde{\chi}^+_2/\tilde{\chi}^-_2 \tilde{\chi}^+_1$ 
pairs the amplitudes are complex
giving rise to a non--zero CP--violating normal chargino
polarization ${\cal P}_N$ with
\begin{eqnarray}
{\cal P}_N [{\tilde{\chi}^-_{1,2}}]
           =\pm 4\lambda^{1/2}\mu_{2,1}\left(F^2_R-F_LF'_L\right)\,
            \sin\Theta\,\sin 2\phi_L\, \sin 2\phi_R\,
            \sin(\beta_L-\beta_R+\gamma_1-\gamma_2)/{\cal N}
\end{eqnarray}

Below, we will concentrate on the production of the lightest charginos.
The direct measurement of chargino
polarization would provide detailed information on the three quartic
charges $Q'_1,Q'_2,Q'_3$. However, the polarization of charginos can only
be determined indirectly from angular distribution of decay
products provided the chargino decay dynamics is known.  
Complementary information can be obtained from the observation of spin--spin
correlations. Since they are reflected in the angular correlations
between the $\tilde{\chi}^-_1$ and $\tilde{\chi}^+_1$ decay products,
some of them are experimentally accessible directly. Moreover, taking 
suitable combinations of polarization and spin--spin correlations, any
dependence on the specific parameters of the chargino decay mechanisms
can be eliminated.\\

The polarization and spin--spin correlations of the charginos are encoded 
in the angular distributions of the decay products.
Assuming the  neutralino $\tilde{\chi}^0_1$ to be the
lightest supersymmetric particle,  
several mechanisms contribute to the decay of the chargino
$\tilde{\chi}^\pm_1$:
\begin{eqnarray*}
\tilde{\chi}^\pm_1 \rightarrow\tilde{\chi}^0_1 +f\bar{f'} \ \ \hskip 1cm
                             [f, f'=l,\nu,q]
\end{eqnarray*}
Choosing the $\tilde{\chi}^\pm_1$ flight direction as quantization axis,
the polar angles of the $f\bar{f'}$ decay systems in the $\tilde{\chi}^-_1/
\tilde{\chi}^+_1$ rest frames are defined as
$\theta^*$ and $\bar{\theta}^*$, respectively, and the corresponding
azimuthal angles with respect to the production plane by
$\phi^*$ and $\bar{\phi}^*$. The spin analysis--powers $\kappa$ and
$\bar{\kappa}$  are the coefficients of those parts of the $\tilde{\chi}^-_1$ 
and $\tilde{\chi}^+_1$ spin--density matrices  which are different from
the unit matrix. The $\kappa$'s are built up by the decay form factors. 
In many scenarios typical numerical values of $\kappa$'s are of the order 
of a few 
10$^{-1}$. For the subsequent analysis they need not be determined in 
detail; it is enough to verify experimentally that they are 
sufficiently large.\\

Integrating out the unobserved production angle $\Theta$ of the charginos
and the invariant masses of the final--state quark or leptonic systems 
$f_i\bar{f}_j$, the differential distribution can be written in terms of
sixteen independent angular parts:
\begin{eqnarray}
\frac{{\rm d}\sigma}{{\rm d}\cos\theta^* {\rm d}\phi^*
               {\rm d}\cos\bar{\theta}^* {\rm d}\bar{\phi}^*} 
&\sim& {\mit \Sigma}_{\rm unpol}\nonumber\\[-2mm]
 &&   +\cos\theta^*\kappa\,{\cal P}
      +\cos\bar{\theta}^*\bar{\kappa}\,\bar{\cal P}\nonumber\\
 &&   +\cos\theta^*\cos\bar{\theta}^*\kappa\bar{\kappa}\,{\cal Q}\nonumber\\
 &&   +\sin\theta^*\sin\bar{\theta}^*\cos(\phi^*+\bar{\phi}^*)
       \kappa\bar{\kappa}\,{\cal Y} + \dots
\label{eq:explicit}
\end{eqnarray}
${\mit \Sigma}_{\rm unpol}$ is the integrated cross section summed over 
chargino polarizations; it can be expressed  in terms of the 
quartic charges $Q_1,Q_2,Q_3$ in analogy to eq.(\ref{eq:initial}):
\begin{eqnarray}
{\mit \Sigma}_{\rm unpol}=4 \int {\rm d}\cos\Theta\,\,
        \left\{(1+\beta^2 \cos^2\Theta)\,Q_1
          +(1-\beta^2)\, Q_2+2\beta\,\cos\Theta\,Q_3\right\}
\end{eqnarray}
where $\beta=\sqrt{1-4m^2_{\tilde{\chi}^\pm_1}/s}$ is the
$\tilde{\chi}^\pm_1$ velocity in the c.m. frame.
Among the polarization vectors, 
only the integrated longitudinal components are useful in the present
context, being proportional to 
\begin{eqnarray}
   {\cal P}= 4\int {\rm d}\cos\Theta\,\left\{(1+\beta^2)\,\cos\Theta\,{Q'}_1
              +4(1-\beta^2)\,\cos\Theta\, {Q'}_2
              +(1+\cos^2\Theta)\beta\, {Q'}_3\right\} 
\end{eqnarray}
for $\tilde{\chi}^-_1$ and $\bar{\cal P}$ for $\tilde{\chi}^+_1$ 
correspondingly. 
The spin correlation ${\cal Q}$ measures the
difference between the cross sections for like--sign and unlike--sign
$\tilde{\chi}^-_1$ and $\tilde{\chi}^+_1$ helicities;
${\cal Y}$ measures the interference between the amplitudes for positive
and negative helicities of both the charginos. They can be expressed in
terms of the quartic charges $Q_1$ to $Q_3$ as
\begin{eqnarray}
&& {\cal Q}=-4\int {\rm d}\cos\Theta\left[(\beta^2+\cos^2\Theta)\,Q_1
         +(1-\beta^2)\cos^2\Theta\, Q_2+2\beta\cos\Theta\, Q_3\right]
         \nonumber \\
&& {\cal Y}=-2\int {\rm d}\cos\Theta(1-\beta^2)\left[Q_1+Q_2\right]\sin^2
  \Theta 
\end{eqnarray}
The terms introduced explicitly in eq.(\ref{eq:explicit}) are
particularly interesting as they can be measured directly in terms of
laboratory observables as follows.\\

The decay angles $\{\theta^*,\phi^*\}$ and
$\{\bar{\theta}^*,\bar{\phi}^*\}$, which are used to measure the
$\tilde{\chi}_1^\pm$ chiralities, are defined in the rest frame of the
charginos $\tilde{\chi}^-_1$ and $\tilde{\chi}^+_1$,
respectively. Since two invisible neutralinos are present in the final
state, they cannot be reconstructed completely.  However, the
longitudinal components and the inner product of the transverse
components can be reconstructed\footnote{The neutralino mass which enters
this analysis, can be predetermined in a model--independent way from
the endpoints of the chargino decay spectra.} from the momenta and 
energies measured in the laboratory
frame (see e.g. Refs.~\cite{KW,A1}),
\begin{eqnarray}
&&\cos\theta^*=\frac{1}{\beta |\vec{p^*}|} \left(\frac{E}{\gamma}-E^*\right)
\ \ , \ \  \cos\bar{\theta}^*=\frac{1}{\beta |\vec{\bar{p^*}}|}
               \left(\frac{\bar{E}}{\gamma}-\bar{E}^*\right) \nonumber\\
&&\sin\theta^*\sin\bar{\theta}^*\cos(\phi^*+\bar{\phi}^*)=
         \frac{|\vec{p}||\vec{\bar{p}}|}{|\vec{p^*}||\vec{\bar{p}^*}|}
         \cos\vartheta+\frac{\left(E-E^*/\gamma \right)
           \left(\bar{E}-\bar{E}^*/\gamma\right)}{\beta^2
           |\vec{p^*}||\vec{\bar{p}^*}|}
\end{eqnarray}
where $\gamma=\sqrt{s}/{2m_{\tilde{\chi}^\pm_1}}$. $E (\bar{E})$ and 
$E^*(\bar{E}^*)$ are the energies of the two hadronic systems in
the $\tilde{\chi}^+_1$ and $\tilde{\chi}^-_1$ decays, defined
in the laboratory frame and in the rest frame of the charginos, respectively; 
$\vec{p} (\vec{\bar{p}})$ and $\vec{p}^* (\vec{\bar{p}^*})$ are 
the corresponding momenta. $\vartheta$ is the angle between the momenta 
of the two hadronic systems in the laboratory frame; 
the angle between the vectors in the transverse plane is given by 
$\Delta \phi^* = 2\pi-(\phi^*+\bar{\phi}^*)$ for the reference frames 
defined earlier. The terms in eq.(\ref{eq:explicit}) can therefore be
measured directly. 
The observables ${\cal P}$, $\bar{\cal P}$, ${\cal Q}$ and ${\cal Y}$
enter into the cross section together with the spin analysis-power 
$\kappa (\bar{\kappa})$. CP--invariance leads to the relation
$\bar{\kappa}=-\kappa$. Therefore, taking the ratios 
${\cal P}\bar{\cal P}/{\cal Q}$ and ${\cal P}\bar{\cal P}/{\cal Y}$, 
these unknown quantities can be eliminated
so that the two ratios reflect unambiguously the properties of the chargino
system, not affected by the neutralinos. It is thus possible to study 
the chargino sector in isolation by measuring the mass of the lightest
chargino, the total production cross section and the spin(--spin) 
correlations.\\

Since the polarization ${\cal P}$ is odd under parity and charge--conjugation, 
it is necessary to identify the chargino electric charges in this case. 
This can be accomplished by making use of the mixed 
leptonic and hadronic decays of the chargino pairs. On the other hand, the
observables ${\cal Q}$ and 
${\cal Y}$ are defined without charge identification so that the 
dominant hadronic decay modes of the charginos can be exploited. \\

\section{MASSES, MIXING ANGLES AND COUPLINGS}

Before the strategies for measuring the masses, mixing angles and the 
couplings are presented in detail, a few general remarks on the structure 
of the chargino system may render the techniques more transparent.\\

\noindent
{\bf (i)} The right--handed cross sections $\sigma_R$ do not involve the 
    exchange of the sneutrino. They depend only, in symmetric form, on the 
    mixing parameters $\cos 2\phi_L$ and $\cos 2\phi_R$.\\

\noindent
{\bf (ii)} The left--handed cross sections $\sigma_L$ and the transverse cross
     section $\sigma_T$ depend on $\cos 2\phi_{L,R}$, the sneutrino
     mass and the $e\tilde{\nu}\tilde{W}$ Yukawa coupling. Thus the sneutrino 
     mass and the Yukawa coupling can be determined from the left-handed and 
     transverse cross sections. 
     [If the sneutrino mass is much larger than the collider energy, 
     only the ratio of the Yukawa coupling over the sneutrino mass squared
     ($g^2_{\tilde{W}}/m^2_{\tilde\nu}$) can be measured by this 
     method \cite{gmpsnu}.]\\

The cross sections $\sigma_L$, $\sigma_R$ and $\sigma_T$ are binomials
in the [$\cos2\phi_L,\cos2\phi_R$] plane. If the two--chargino model is 
realized in nature, any two contours, $\sigma_L$ and $\sigma_R$ for example, 
will at least cross at one point in the plane between $-1 \leq \cos2\phi_L,
\cos2\phi_R \leq +1$.  However, the contours, being ellipses or hyperbolae, 
may cross 
up to four times. This ambiguity can be resolved by measuring the third 
physical quantity, $\sigma_T$ for example. 
The measurement of $\sigma_T$ is particularly important if the sneutrino
mass is unknown. While the curve for $\sigma_R$ is fixed, the curve
for $\sigma_L$ will move in the $[\cos2\phi_L,\cos2\phi_R]$ plane with 
changing $m_{\tilde\nu}$. However, the third curve will intersect the other 
two in the same point only if the mixing angles as well as the sneutrino 
mass correspond to the correct physical values.\\

The numerical analyses presented below  have been worked out for the two 
parameter points introduced in Ref.\cite{LCWS}.     
They correspond to a
small and a large $\tan\beta$ solution for universal gaugino and
scalar masses at the GUT scale:
\begin{eqnarray}
&&\mbox{\boldmath $RR1$}:
  (\tan\beta, m_0,M_{\frac{1}{2}})=(\,\,3,100\,{\rm GeV}, 200\, {\rm GeV})
  \nonumber\\
&&\mbox{\boldmath $RR2$}:
  (\tan\beta, m_0,M_{\frac{1}{2}})=(30, 160\,{\rm GeV}, 200\, {\rm GeV})
\label{eq:parameter}
\end{eqnarray}
The CP-phase $\Phi_\mu$ is set to zero. The induced chargino
$\tilde{\chi}^\pm_{1,2}$, neutralino $\tilde{\chi}^0_1$ and sneutrino
$\tilde{\nu}$ masses are given as follows:
\begin{eqnarray}
&& m_{\tilde{\chi}^\pm_1}=128/132\,{\rm GeV}\,  \qquad
   m_{\tilde{\chi}^0_1}\,  = 70/72\,{\rm GeV} 
\nonumber\\
&& m_{\tilde{\chi}^\pm_2}=346/295\,{\rm GeV}\, \qquad
   m_{\tilde{\nu}}       =166/206 \,{\rm GeV}
\label{eq:measured masses}
\end{eqnarray}
for the two points \mbox{\boldmath $RR1/2$}, respectively.
The size of the unpolarized total cross sections
$\sigma[e^+e^- \rightarrow \tilde\chi_i^+ \tilde\chi_j^-]$ as functions 
of the collider energy is shown for two reference points in Fig.~2. 
With the 
maximum of the cross sections in the range of 0.1 to 0.3~pb, about 
$10^5$ to $3\times10^5$ events can be generated for an integrated luminosity
$\int {\cal L} \simeq 1\,\, {\rm ab}^{-1}$ as planned in three years 
of running at TESLA.\\

The cross sections for chargino pair--production  rise steeply at the 
threshold,
\begin{eqnarray}
\sigma[e^+e^- \rightarrow \tilde\chi_i^+ \tilde\chi^-_j] \sim
\sqrt{s-(m_{\tilde{\chi}^\pm_i}+m_{\tilde{\chi}^\pm_j})^2}
\end{eqnarray}
so that the masses $m_{\tilde\chi^\pm_1}$, $m_{\tilde\chi^\pm_2}$
can be measured very accurately in the production processes of the
final--state pairs \{11\}, \{12\} and \{22\}. 
Detailed experimental simulations 
have shown that 
accuracies $\Delta m_{\tilde\chi_1^\pm}$ =~40~MeV and 
$\Delta m_{\tilde\chi_2^\pm}$=250~MeV can be achieved in 
high--luminosity threshold scans \cite{martyn}.

\bigskip

\subsection{Light chargino pair production}

At an early phase of the $e^+e^-$ linear collider the energy may only be
sufficient to reach the threshold of the light chargino pair
$\tilde{\chi}^+_1\tilde{\chi}^-_1$. Nearly the entire structure of the 
chargino system can nevertheless be reconstructed even in this case.\\

\subsubsection{Exploiting longitudinal and transverse beam polarization}

By analyzing the $\{11\}$ mode in $\sigma_L\{11\}$, $\sigma_R\{11\}$, the
mixing angles $\cos2\phi_L$ and $\cos2\phi_R$ can be determined up to
at most a four--fold ambiguity if the sneutrino mass is known 
and the Yukawa coupling is identified with the gauge coupling. 
The ambiguity can be resolved by adding the information from 
$\sigma_T\{11\}$. This is demonstrated\footnote{With event numbers of
order $10^5$, statistical errors are at the per--mille level.} in Fig.~3 
for the reference point \mbox{\boldmath $RR1$}
at the energy $\sqrt{s}=400$ GeV. 
Moreover, the additional measurement of the
transverse cross section can be exploited to determine the
sneutrino mass. While the right--handed cross section $\sigma_R$ 
does not depend on $m_{\tilde \nu_e}$, the contours $\sigma_L$, $\sigma_T$ 
move uncorrelated in the $[\cos2\phi_L,\cos2\phi_R]$ plane until the correct 
sneutrino mass is used in the analysis. The three contour lines intersect 
exactly in one point of the plane only if all the parameters correspond to the
correct physical values.\\

\subsubsection{Chargino polarization}

Without longitudinal and transverse beam polarizations, the polarization
of the charginos in the final state and their spin--spin correlations
can be used to determine
the mixing angles $\cos 2\phi_L$ and $\cos 2\phi_R$.\\

The observables ${\cal P}$, $\bar{\cal P}$, ${\cal Q}$ and ${\cal Y}$
enter into the cross section together with the spin analysis-power 
where $\bar{\kappa}=-\kappa$ in CP--invariant theories. 
Therefore, taking the ratios ${\cal P}^2/{\cal Q}$
and ${\cal P}^2/{\cal Y}$, these unknown quantities can be eliminated
so that the two ratios reflect unambiguously the properties of the chargino
system, not affected by the neutralinos. It is thus possible to study 
the chargino sector in isolation by measuring the mass of the lightest
chargino, the total production cross section and the spin(--spin) correlations.
The energy dependence of the two ratios ${\cal P}^2/{\cal Q}$ and
${\cal P}^2/{\cal Y}$ is shown in Fig.~4; the same parameters are chosen
as in the previous figures. The two ratios are sensitive to the quartic charges
at sufficiently large c.m. energies since the charginos are, on the average,
unpolarized at the threshold, c.f. eq.~(26).
Note that ${\cal Y}$ vanishes for asymptotic energies so that an optimal
energy must be chosen not far above threshold to measure this observable.\\

The measurement of the cross section at an energy $\sqrt{s}$, 
and either of the ratios ${\cal P}^2/{\cal Q}$ or ${\cal P}^2/{\cal Y}$ 
can be interpreted as 
contour lines in the plane $[\cos 2\phi_L,\cos 2\phi_R]$ 
which intersect at large angles so that a high precision in the 
resolution can be achieved. A representative example for the 
determination of $\cos 2\phi_L$ and $\cos 2\phi_R$ is shown in 
Fig.~5 for the reference point \mbox{\boldmath{$RR1$}}. 
The mass of the light chargino is set to $m_{\tilde{\chi}^\pm_1}=128$ 
GeV, and the ``measured'' cross section, ${\cal P}^2/{\cal Q}$ and 
${\cal P}^2/{\cal Y}$  are taken to be 
\begin{eqnarray}
\sigma\{11\}=0.32\,\, {\rm pb}\,,\quad
{\cal P}^2/{\cal Q}=-0.63\,,\quad 
{\cal P}^2/{\cal Y}=-6.46
\label{eq:measure}
\end{eqnarray}
at the $e^+e^-$ c.m. energy $\sqrt{s}=400$ GeV.
The three contour lines meet at a single point $[\cos 2\phi_L,\cos 2\phi_R]
=[0.645,0.844]$.\\

\subsection{ The complete chargino system}

From the analysis of the complete chargino system 
$\{\tilde\chi_1^+ \tilde\chi_1^-, \tilde\chi_1^+ \tilde\chi_2^-,
\tilde\chi_2^+ \tilde\chi_2^-\}$, together with the knowledge of the 
sneutrino mass from sneutrino pair production, the maximal information 
can be extracted on the basic parameters of the electroweak SU(2) gaugino 
sector. Moreover, the identity of the $e \tilde\nu \tilde W$ Yukawa 
coupling with the $e \nu W$ gauge coupling, which is of fundamental nature
in supersymmetric theories, can be tested very accurately. This 
analysis is the final target of LC experiments which should provide
a complete picture of the electroweak gaugino sector with resolution at 
least at the per-cent level.\\

The case will be exemplified for the scenario 
\mbox{\boldmath $RR1$} with $\tan\beta=3$ while the final results will
also be presented for \mbox{\boldmath $RR2$} with $\tan\beta=30$. 
To simplify the picture, without loss of generality, we will not choose 
separate energies at the maximal values of the cross sections, but instead 
we will work with a single collider energy $\sqrt{s}$=
800 GeV and an integrated luminosity $\int {\cal L} =1\,\, {\rm ab}^{-1}$. 
The polarized cross sections take the following values:
\begin{eqnarray}
\begin{array}{lll}
 \sigma_R\{11\}=\,\,1.8\,{\rm fb}\, &{ }\hskip 3mm
 \sigma_L\{11\}=787.7\,{\rm fb}\,   &{ }\hskip 3mm
 \sigma_T\{11\}=0.53\,{\rm fb}  \\
 \sigma_R\{12\}=12.1\,{\rm fb}\, &{ }\hskip 3mm
 \sigma_L\{12\}=106.2\,{\rm fb}\,&{ }\hskip 3mm
 \sigma_T\{12\}=0.53\,{\rm fb} \\
 \sigma_R\{22\}=67.1\,{\rm fb}\, &{ }\hskip 3mm
 \sigma_L\{22\}=337.5\,{\rm fb}\, & { }\hskip 3mm
 \sigma_T\{22\}=1.07\,{\rm fb}
\end{array}
\label{eq:measured x-section}
\end{eqnarray}

Chargino pair production with right-handed electron beams provides us 
with the cross sections $\sigma_{R_i}$($i=\{11\},\{12\},\{22\}$). Due to the 
absence of the sneutrino exchange diagram, the cross sections can be 
expressed symmetrically in the mixing parameters
\begin{eqnarray}
&& c_{2L}=\cos2\phi_L\nonumber\\
&& c_{2R}=\cos2\phi_R
\end{eqnarray}
as follows:
\begin{eqnarray}
\sigma_{R_i} = A_{R_i}\,(c^2_{2L}+c^2_{2R})+B_{R_i}\,(c_{2L}+c_{2R})
+C_{R_i}\, c_{2L}c_{2R}+D_{R_i}\ \ \left(i=\{11\},\{12\},\{22\}\right)
\label{eq:sigmar}
\end{eqnarray}
The coefficients $A_{R_i}$, $B_{R_i}$, $C_{R_i}$ and $D_{R_i}$  
involve only known parameters, the chargino masses and the
energy. Depending on
whether $A^2_{R_i} \stackrel{>}{{ }_<} C^2_{R_i}/4$, the contour
lines for $\sigma_R\{11\}, \sigma_R\{12\}$, $\sigma_R\{22\}$
in the [$c_{2L},c_{2R}$] plane (cf. Fig.6) are either 
closed ellipses or open hyperbolae\footnote{The cross
section $\sigma_R\{12\}$ is always represented by an ellipse.}. 
They intersect in two points of the plane which are symmetric under the 
interchange $c_{2L}
\leftrightarrow c_{2R}$; for \mbox{\boldmath $RR1$}: [$c_{2L},c_{2R}$]
=[0.645,0.844]
and interchanged.\\ 

While the right--handed cross sections do not involve sneutrino exchange,
the cross sections for left--handed electron beams are dominated 
by the sneutrino contributions unless the sneutrino mass is very large. 
In general, the three observables $\sigma_{L_i}$ ($i=\{11\},\{12\},\{22\}$) 
exhibit quite a different dependence on $c_{2L}$ and  
$c_{2R}$. In particular, they are not symmetric with respect 
to $c_{2L}$ and $c_{2R}$ so that   
the correct solution for $[c_{2L},c_{2R}]$ can be 
singled out of the two solutions obtained from the right-handed cross 
sections eq.(\ref{eq:sigmar}). As before, the three observables 
can be expressed as   
\begin{eqnarray}
\sigma_{L_i}= A_{L_i}\,c^2_{2L}+A'_{L_i}\,c^2_{2R}+ B_{L_i}\,c_{2L}
           +B'_{L_i}\,c_{2R}+C'_{L_i}\,c_{2L}c_{2R}+D'_{L_i}\ \ 
           \left(i=\{11\},\{12\},\{22\}\right)
\label{eq:sigmal}
\end{eqnarray}
The coefficients of the linear and quadratic terms of $c_{2L}$ and 
$c_{2R}$ depend on known parameters only. The shape of the contour lines
is given by the chargino masses and the sneutrino mass, being either elliptic 
or hyperbolic for $A_{L_i} A'_{L_i} \stackrel{>}{{ }_<} C^2_{L_i}/4$, 
respectively. These asymmetric equations are satisfied {\it only} by one 
solution, as shown in Fig.~6. Among the two 
solutions obtained above from $\sigma_{R_i}$ only the set 
$[c_{2L},c_{2R}]=[0.645,0.844]$ 
satisfies eq.(\ref{eq:sigmal}).\\

At the same time, the identity between the $e \tilde\nu \tilde W$
Yukawa coupling and the $e \nu W$ gauge coupling can be tested.
Varying 
the Yukawa coupling freely, the contour lines $\sigma_{L_i}$ move
through the [$c_{2L},c_{2R}$] plane. Only for the supersymmetric 
solutions the curves $\sigma_{L_i}$ intersect each other and the curves
$\sigma_{R_i}$ in exactly one point. Combining the analyses of 
$\sigma_{R_i}$ and $\sigma_{L_i}$, the masses, the mixing
parameters and the Yukawa coupling can be determined to quite a high 
precision\footnote{In contrast to the restricted
$\tilde{\chi}^+_1\tilde{\chi}^-_1$ case, it is not 
necessary to use transversely polarized beams to determine this set of 
parameters unambiguously. If done so nevertheless, the analysis follows the 
same steps as discussed above. The additional information will reduce the 
errors on the fundamental parameters.}
\begin{eqnarray}
&&{ }\hskip -1.2cm m_{\tilde\chi_1^\pm}=128\pm 0.04\, {\rm GeV}\quad 
   \cos2\phi_L=0.645 \pm 0.02  \qquad
   g_{\tilde{W}}/g_W= 1\pm 0.001 \nonumber\\  
&&{ }\hskip -1.2cm m_{\tilde\chi_2^\pm}=346\pm 0.25\, {\rm GeV}\quad
   \cos2\phi_R=0.844 \pm 0.005 
\label{eq:measured}
\end{eqnarray}
The 1$\sigma$ statistical errors have been derived for an integrated 
luminosity of $\int {\cal L} =1\,\,{\rm ab}^{-1}$.\\

Thus the parameters of the chargino system, masses $m_{\tilde\chi_1^\pm}$
and  $m_{\tilde\chi_2^\pm}$, mixing parameters $\cos2\phi_L$ and 
$\cos2\phi_L$, as well as the Yukawa coupling can be used to extract the 
fundamental parameters of the underlying supersymmetric theory
with high accuracy.

\section{THE FUNDAMENTAL SUSY PARAMETERS}
\label{fundamental}

\subsection{The $\tilde{\chi}^\pm_1$ base}

From the analysis of the $\tilde{\chi}^\pm_1$ states alone, the mixing 
parameters $\cos 2\phi_L$ and $\cos 2\phi_R$ can be derived unambiguously.
This information is sufficient to derive the fundamental gaugino 
parameters $\{M_2,\mu,\tan\beta\}$ in CP--invariant theories up to at 
most a discrete two--fold ambiguity. \\

The solutions can be discussed most transparently by introducing the two
triangular quantities
\begin{eqnarray}
p\backslash q=\cot(\phi_R\mp\phi_L)
\end{eqnarray}
These two quantities can be expressed in terms of the mixing angles:
\begin{eqnarray}
&& p=\pm\left|\frac{\sin 2\phi_L+\sin 2\phi_R}{
                                        \cos 2\phi_L-\cos 2\phi_R}
                             \right|\nonumber\\
&& q=\frac{1}{p}\,\frac{\cos 2\phi_L+\cos 2\phi_R}{
                                            \cos 2\phi_L-\cos 2\phi_R}
\end{eqnarray}
Apart from the overall sign ambiguity of the pair $(p,q)$ which can
be removed by definition, the set is 
two--fold ambiguous due to the unfixed relative sign between $\sin 2\phi_L$ 
and $\sin 2\phi_R$.\\

From the solutions $(p,q)$ derived above, the SUSY parameters can be
determined in the following way:
\begin{eqnarray}
\cos 2\phi_R \stackrel{>}{{ }_<} \cos 2\phi_L\ \ : \ \ 
\tan\backslash\cot\beta = \frac{p^2-q^2\pm 2\sqrt{\chi^2(p^2+q^2+2-\chi^2)}}{
       (\sqrt{1+p^2}-\sqrt{1+q^2})^2-2\chi^2} 
      \; \Rightarrow \; \tan\beta\stackrel{>}{{ }_<} 1
\end{eqnarray}
where $\chi^2 =m^2_{\tilde{\chi}^\pm_1}/m^2_W$.
The gaugino and higgsino mass parameters are given in terms of $p$ and 
$q$ by 
\begin{eqnarray}
M_2&=&\frac{m_W}{\sqrt{2}}\bigg[(p+q)\sin\beta-(p-q)\cos\beta\bigg]
       \nonumber\\
\mu&=&\frac{m_W}{\sqrt{2}}\bigg[(p-q)\sin\beta-(p+q)\cos\beta\bigg]
\label{eq:M2_mu}
\end{eqnarray}
The parameters $M_2$, $\mu$ are uniquely fixed if $\tan\beta$ is chosen 
properly. Since $\tan\beta$ is invariant
under pairwise reflection of the signs in $(p,q)$, the definition 
$M_2 > 0$ can be exploited to remove this additional ambiguity.\\

As a result, the fundamental SUSY parameters $\{M_2,\mu, \tan\beta\}$ 
can be derived from the observables $m_{\tilde{\chi}^\pm_1}$
and $\cos 2\phi_R$, $\cos 2\phi_L$ up to at most a two--fold ambiguity.\\

\subsection{The complete set of the fundamental SUSY parameters}

From the set $m_{\tilde{\chi}^\pm_{1,2}}$ and $\cos 2\phi_{L,R}$ of 
measured observables, the fundamental supersymmetric parameters 
$\{M_2, |\mu|, \cos\Phi_\mu, \tan\beta\}$ in CP--(non)invariant theories 
can be determined unambiguously in the following way.\\

\noindent
{\bf (i) \boldmath{$M_2,|\mu|$}}: Based on  the definition 
$M_2>0$, the gaugino mass parameter $M_2$ and the modulus of the higgsino
mass parameter read as follows:
\begin{eqnarray}
M_2&=&\sqrt{(m_{\tilde{\chi}^\pm_2}^2+m_{\tilde{\chi}^\pm_1}^2
               -2m^2_W)/2
         -(m_{\tilde{\chi}^\pm_2}^2-m_{\tilde{\chi}^\pm_1}^2)
          (\cos 2\phi_L+\cos 2\phi_R)/4}\nonumber\\
|\mu|&=&\sqrt{(m_{\tilde{\chi}^\pm_2}^2+m_{\tilde{\chi}^\pm_1}^2
               -2m^2_W)/2
         +(m_{\tilde{\chi}^\pm_2}^2-m_{\tilde{\chi}^\pm_1}^2)
          (\cos 2\phi_L+\cos 2\phi_R)/4}
\label{eq:M2mu}
\end{eqnarray} 

\noindent
{\bf (ii) $\mbox{\boldmath{$\cos\Phi$}}_\mu$}:
The sign of $\mu$ in CP--invariant theories and, more generally, the cosine 
of the phase of $\mu$ in CP--noninvariant theories is determined by 
the $\tilde{\chi}^\pm_1,\tilde{\chi}^\pm_2$ masses and $\cos 2\phi_{L,R}
=c_{2L,R}$:
{\scriptsize
\begin{eqnarray*}
\cos\Phi_\mu=\frac{(m_{\tilde{\chi}^\pm_2}^2-m_{\tilde{\chi}^\pm_1}^2)^2
                   (2-c^2_{2L}-c^2_{2R})
                  -8 m^2_W(m_{\tilde{\chi}^\pm_2}^2
                             +m_{\tilde{\chi}^\pm_1}^2-2m^2_W)}{
         \sqrt{\left[16m^4_W
               -(m_{\tilde{\chi}^\pm_2}^2-m_{\tilde{\chi}^\pm_1})^2
               (c_{2L}-c_{2R})^2\right]
               \left[4 (m_{\tilde{\chi}^\pm_2}^2+m_{\tilde{\chi}^\pm_1}^2
               -2m^2_W)^2-(m_{\tilde{\chi}^\pm_2}^2-m_{\tilde{\chi}^\pm_1}^2)^2
               (c_{2L}+c_{2R})^2\right]}}
\label{eq:phi_mu}
\end{eqnarray*}
} \\[-2.55cm]
\begin{eqnarray}
\mbox{ }
\end{eqnarray}\\[2mm]
\noindent
{\bf (iii) \boldmath{$\tan\beta$}}: The value of $\tan\beta$ 
is uniquely determined in terms of two chargino masses and two mixing 
angles:
\begin{eqnarray}
\tan\beta=\sqrt{\frac{4m^2_W
               -(m_{\tilde{\chi}^\pm_2}^2-m_{\tilde{\chi}^\pm_1}^2)
               (\cos 2\phi_L-\cos 2\phi_R)}{
                      4m^2_W
               +(m_{\tilde{\chi}^\pm_2}^2-m_{\tilde{\chi}^\pm_1}^2)
               (\cos 2\phi_L-\cos 2\phi_R)}}\,
\label{eq:tanb}
\end{eqnarray}
As a result, the fundamental SUSY parameters $\{
M_2,\mu, \tan\beta\}$ in CP--invariant
theories, and $\{M_2,\, |\mu|,\, \cos\Phi_\mu,\\ \tan\beta\}$
in CP--noninvariant theories, 
can be extracted {\rm unambiguously} from the observables 
$m_{\tilde{\chi}^\pm_{1,2}}$, $\cos 2\phi_R$, and $\cos 2\phi_L$.
The final ambiguity in $\Phi_\mu \leftrightarrow 2 \pi - \Phi_\mu$ 
in CP--noninvariant theories must be resolved by measuring observables
related to the 
normal $\tilde{\chi}^-_1$ or/and $\tilde{\chi}^+_2$  polarization in 
non--diagonal $\tilde{\chi}^-_1\tilde{\chi}^+_2$ chargino--pair 
production \cite{R11}.\\
 
For illustration, the accuracy which can be expected in such an analysis,
is shown for both CP--invariant reference points \mbox{\boldmath $RR1$} 
and \mbox{\boldmath $RR2$} 
in Table 2.  If $\tan\beta$ is large, this parameter is difficult to
extract from the chargino sector. Since the chargino observables depend 
only on $\cos2\beta$, the dependence on $\beta$ is flat for 
$2\beta\rightarrow \pi$ so that eq.(\ref{eq:tanb}) is not very 
useful to derive the value of
$\tan\beta$ due to error propagation. A significant lower bound can
be derived nevertheless in any case. 

\begin{table}[\hbt]
\caption[{\bf Table 2:}]{\label{tab:measured}
{\it Estimate of the accuracy with which the parameters $M_2$, $\mu$,
     $\tan\beta$ can be determined, including sgn($\mu$),
     from chargino masses and production cross sections; errors at the
     1$\sigma$ level are statistical only.}}
\begin{center}
\begin{tabular}{|c|c|c|c|c|}\hline
    & \multicolumn{2}{|c|}{\mbox{  }} & 
      \multicolumn{2}{c|}{\mbox{  }}\\[-4mm]
    & \multicolumn{2}{|c|}{\mbox{\boldmath $RR1$}} & 
      \multicolumn{2}{c|}{\mbox{\boldmath $RR2$}}\\
  \cline{2-5}
& theor. value & fit value & theor. value & fit value \\ 
  \cline{2-5} \hline 
  &  &  &  & \\[-3mm] 
$M_2$      & 152\, GeV & $152\pm 1.75$\, GeV & 150\, GeV & $150\pm 1.2$\, GeV \\
$\mu$      & 316\, GeV & $316\pm 0.87$\, GeV & 263\, GeV & $263\pm 0.7$\, GeV \\
$\tan\beta$& 3         & $3\pm 0.69$         & 30        &  $> 20.2$   \\[1mm]
\hline
\end{tabular}
\end{center}
\end{table}
%

\subsection{ Two--state completeness relations}

The two--state mixing of charginos leads to sum rules for the chargino
couplings. They can be formulated in terms of the squares of the
bilinear charges, {\it i.e.} the elements of the quartic charges.
This follows from the observation that the mixing matrix is
built up by trigonometric functions among which many relations are
valid.  From evaluating these sum rules experimentally, it can be concluded 
whether the two--chargino system $\{\tilde\chi_1^\pm, \tilde\chi_2^\pm\}$ 
forms a closed system, or whether additional states, at high mass scales, 
mix in.\\

The following general sum rules can be derived for the two--state 
charginos system at tree level:
\begin{eqnarray}
\sum_{i,j=1,2}|Q_{\alpha\beta}|^2\{ij\} = 2\,(|D_\alpha|^2+|F_\alpha|^2)
\qquad (\alpha\beta)=(LL,RL,RR) 
\label{eq:sum rule}
\end{eqnarray}
\noindent
The right--hand side is independent of any supersymmetric parameters, and
it depends only on the electroweak parameters $\sin^2\theta_W, m_Z$ 
and on the energy, cf. eq.(\ref{eq:DFLR}).  Asymptotically, the initial 
energy dependence and the $m_Z$ dependence drop out. The corresponding 
sum rule for the mixed left--right (LR) combination,
\begin{eqnarray}
\sum_{i,j=1,2}|Q_{LR}|^2\{ij\} = 2(|D'_L|^2+|F'_L|^2)
\end{eqnarray}
involves the sneutrino mass and Yukawa coupling.\\
  
The validity of these sum rules is reflected in both the quartic 
charges and the production cross sections. However, due to mass effects
and the $t$--channel sneutrino exchange, it is not straightforward to derive 
the sum rules for the quartic charges and the production cross sections 
in practice. Only {\it asymptotically} at high energies the sum rules 
(\ref{eq:sum rule})
for the charges can be transformed directly into sum rules for the
associated cross sections:
\begin{eqnarray}
\sum_{i,j=1,2}\sigma_{L,R}\{ij\} \simeq \frac{16\pi\alpha^2}{3s}
             \left(|D_{L,R}|^2+|F_{L,R}|^2\right)
\end{eqnarray}

For non--asymptotic energies the fact that all the physical 
observables are bilinear in $\cos2\phi_L$ and $\cos2\phi_R$, enables 
us nevertheless to relate the cross
sections with the set of the six variables 
$\vec{z}=\{1,c_{2L},\,c_{2R},\,c^2_{2L},\, c^2_{2R},\\ c_{2L}c_{2R}\}$. 
For the sake of simplicity we restrict ourselves to the left and 
right--handed cross sections. We introduce the
generic notation $\vec{\sigma}$ for the
six cross sections $\sigma_R\{ij\}$ and $\sigma_L\{ij\}$:
\begin{eqnarray}
\vec{\sigma} =\bigg\{\sigma_R\{11\},\sigma_R\{12\},\sigma_R\{22\},
                     \sigma_L\{11\},\sigma_L\{12\},\sigma_L\{22\}\bigg\}\,
\end{eqnarray}
Each cross section can be decomposed in terms of $c_{2L}$ and $c_{2R}$ 
by noting that 
\begin{eqnarray}
{\sigma}_i = \sum_{j=1}^6 \, 
    f_{ij}[m^2_{\tilde{\chi}^\pm_{1,2}},m^2_{\tilde{\nu}}] \,z_j
\end{eqnarray}
The matrix elements $f_{ij}$ can easily be derived from Table 1 together 
with eqs.(\ref{eq:DFLR}-\ref{eq:[12]}).
Since the observables $\sigma_R$ 
do not involve sneutrino contributions, the corresponding functions 
$f_{ij}$ do not depend on the sneutrino mass.
The 6$\times$6 matrix $f_{ij}$ relates the six left/right--handed cross 
sections
and the six variables $z_i$. Inverting the matrix gives the expressions 
for the variables $z_i$ in
terms of the observables. Since the variables $z_i$ are not independent, we
obtain several non--trivial relations among the observables
of the chargino sector: 
\begin{eqnarray}
&& z_1=1\ \ \ \  :\ \ f^{-1}_{1j} {\sigma}_j = 1 \\[1mm]
&& z_4=z^2_2 \, \ \ :\ \ f^{-1}_{4j} {\sigma}_j 
         =\left[ f^{-1}_{2j} {\sigma}_j \right]^2 \\[1mm]
&& z_5=z^2_3\, \ \ : \ \ f^{-1}_{5j} {\sigma}_j 
         =\left[ f^{-1}_{3j} {\sigma}_j\right]^2 \\[1mm]
&& z_6=z_2 z_3 :\ \ f^{-1}_{6j} {\sigma}_j 
         =f^{-1}_{2j} f^{-1}_{3k} {\sigma}_j{\sigma}_k\,
\end{eqnarray}
where summing over repeated indices is understood. The failure of
saturating any of these sum rules by the measured cross sections would
signal that the chargino two--state $\{\tilde\chi_1^\pm$,
$\tilde\chi_2^\pm\}$ system is not complete and additional states mix
in.\\

\section{CONCLUSIONS}

We have analyzed in this report how the parameters of the chargino 
system, the chargino masses $m_{\tilde{\chi}^\pm_{1,2}}$
and the size of the wino and higgsino components in the chargino 
wave--functions, parameterized by the two mixing angles 
$\phi_L$ and $\phi_R$, can be extracted from pair production of 
the chargino states in $e^+e^-$ annihilation. 
Three production cross sections $\tilde{\chi}^+_1 \tilde{\chi}^-_1$, 
$\tilde{\chi}^+_1 \tilde{\chi}^-_2$, $\tilde{\chi}^+_2 \tilde{\chi}^-_2$,
for left-- and right--handedly polarized electrons
give rise to six independent observables.
The method is independent of the chargino decay properties, {\it i.e.}
the analysis is not affected by the structure of the neutralino sector
which is generally very complex in supersymmetric theories while the 
chargino sector remains generally isomorphic to the minimal form of the
MSSM.  \\

The measured
chargino masses $m_{\tilde{\chi}^\pm_{1,2}}$ and the 
two mixing angles $\phi_L$ and $\phi_R$ allow us to extract the fundamental 
SUSY parameters $\{M_2,\mu, \tan\beta\}$ in CP--invariant theories
unambiguously; in CP--noninvariant theories the
modulus of $\mu$ and the cosine of the phase can be determined,
leaving us with just a discrete two--fold ambiguity $\phi_\mu \leftrightarrow
2\pi-\phi_\mu$ which can be
resolved by measuring the sign of observables associated with the normal 
$\tilde{\chi}^\pm_{1,2}$ polarizations.\\

Sum rules for the production cross sections can be used at high energies
to check whether the two--state chargino system is a closed system or whether
additional states mix in from potentially high scales.\\

{\it To summarize}, the measurement of the processes
$e^+e^-\rightarrow \tilde{\chi}^+_i \tilde{\chi}^-_j$ [$i,j=1,2$] 
carried out with polarized beams, leads to a complete analysis of the basic
SUSY parameters $\{M_2, \mu, \tan\beta\}$ in the chargino sector.
Since the analysis can be performed with high precision, this set provides
a solid platform for extrapolations to scales eventually near the Planck scale
where the fundamental supersymmetric theory may be defined.

\subsubsection*{Acknowledgments}

This work was supported in part by the Korea Science and Engineering 
Foundation (KOSEF) through the KOSEF-DFG large collaboration project, 
Project No. 96-0702-01-01-2, and in part by the Center for Theoretical 
Physics. JK was supported by the KBN Grant No. 2P03B 052 16 and
MG by the Alexander von Humboldt Foundation Stiftung.

\newpage

\begin{figure}
 \begin{center}
\epsfig{figure=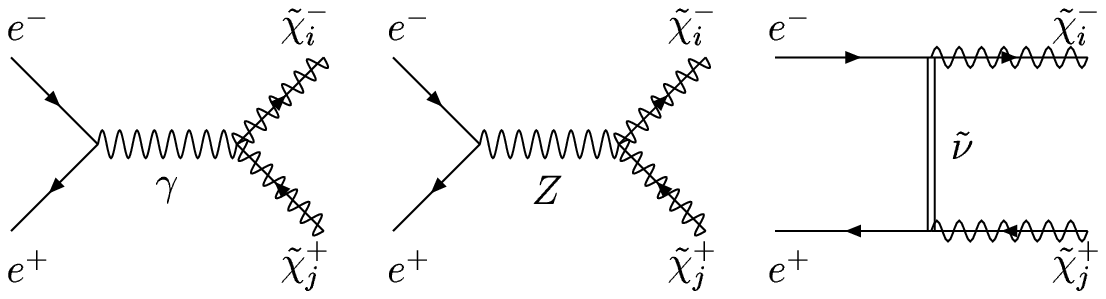,width=16cm,height=7cm}
 \end{center}
\vskip -1cm
\caption[{\bf Figure 1:}]
         {\it The three mechanisms contributing to the production
         of chargino  pairs $\tilde{\chi}^-_i \tilde{\chi}^+_j$ in 
         $e^+e^-$ collisions.}
\label{fig1}
\end{figure}

\newpage
\begin{figure}
 \begin{center}
\hbox to\textwidth{\hss\epsfig{file=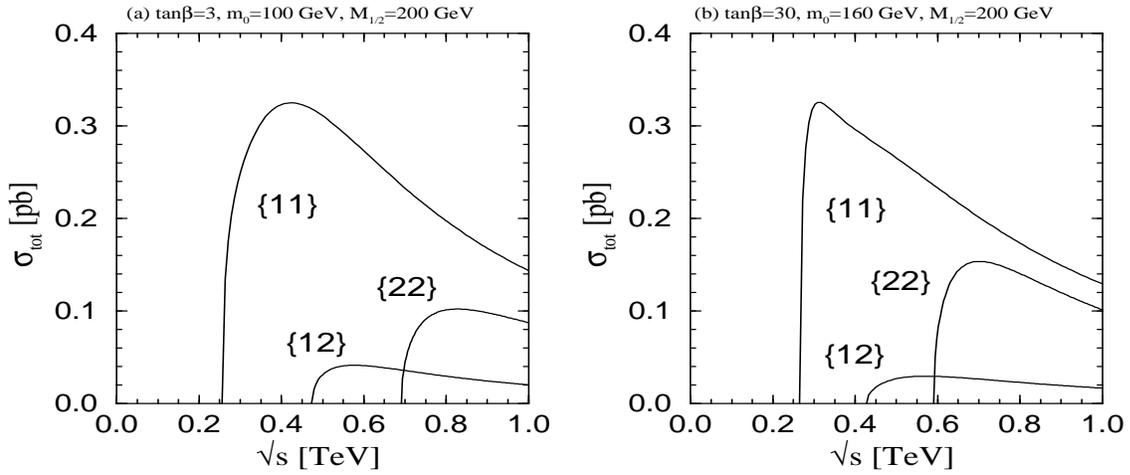,width=15cm,height=6.5cm}\hss}
 \end{center}
\vskip -0.8cm
\caption[{\bf Figure 2:}]
         {\it The cross sections for the production of charginos as a
         function of
         the c.m. energy (a) with the $\mbox{\boldmath $RR1$}$ set and
         (b) with the $\mbox{\boldmath $RR2$}$ set of the fundamental
         SUSY parameters.}
\label{fig2}
\end{figure}

\vskip 0.5cm 
\vskip -0.6cm
\begin{figure}
 \begin{center}
  \begin{picture}(140,180)
   \put(-85,0){\epsfig{file=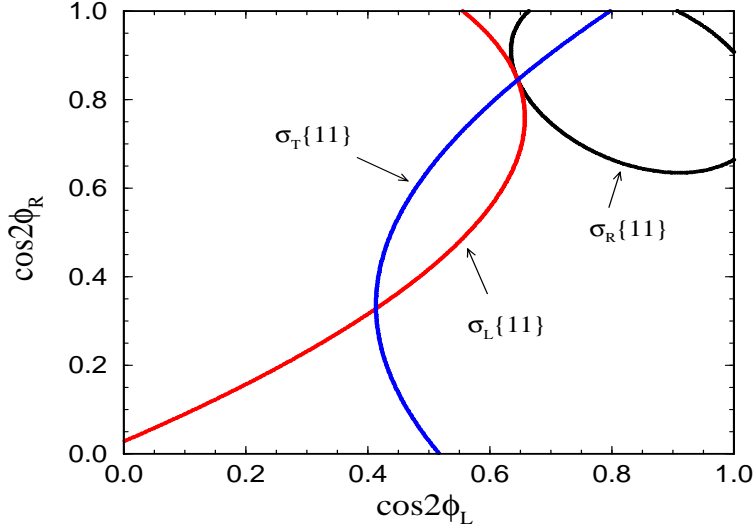,width=10cm,height=7.2cm}}
  \end{picture}
 \end{center}
 \vskip -0.6cm
\caption[{\bf Figure 3:}]
        {\it Contours of the cross sections $\sigma_L\{11\}$,~$\sigma_R\{11\}$ 
         and $\sigma_T\{11\}$ in the $[\cos2\phi_L,\cos2\phi_R]$ plane 
         for the set \mbox{\boldmath $RR1$}
         $[\tan\beta=3,~m_0=100~{\rm GeV},
         ~M_{1/2}=200~{\rm GeV}]$ at the $e^+e^-$ c.m. energy of 400 GeV.} 
\label{fig3}
\end{figure}

\vskip 0.2cm
\begin{figure}[htb]
\begin{center}
\hbox to\textwidth{\hss\epsfig{file=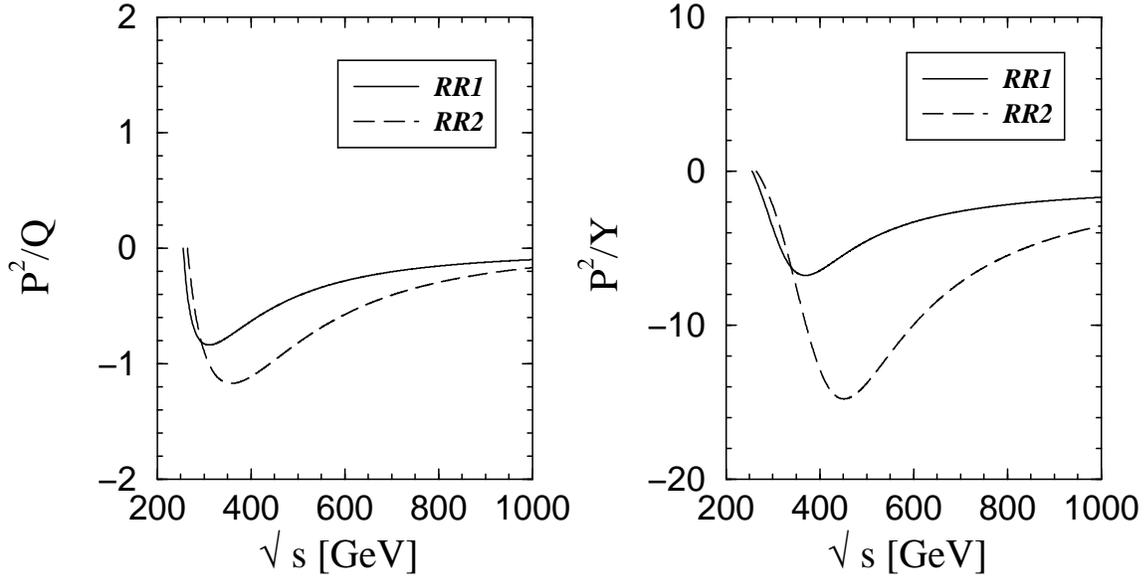,width=15cm,height=8cm}\hss}
\caption[{\bf Figure 4:}]
        {\it The energy dependence of the ratios ${\cal P}^2/{\cal Q}$
             and ${\cal P}^2/{\cal Y}$: solid line for the set 
             \mbox{\boldmath $RR1$} $[\tan\beta=3,~m_0=100~{\rm GeV}, 
             ~M_{1/2}=200~{\rm GeV}]$ and dashed line for the set
             \mbox{\boldmath $RR2$} $[\tan\beta=30,~m_0=160~{\rm GeV}, 
             ~M_{1/2}=200~{\rm GeV}]$.}
\label{fig4}
\end{center}
\end{figure}

\newpage
\voffset -2cm
%
\begin{figure}[htb]
\begin{center}
  \begin{picture}(140,180)
   \put(-85,0){\epsfig{file=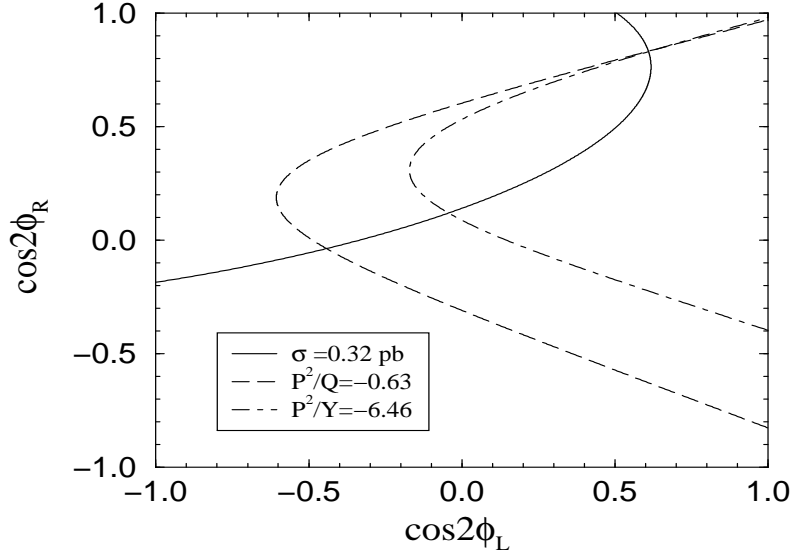,width=10.5cm,height=9cm}}
  \end{picture}
\vskip -0.5cm
\caption[{\bf Figure 5:}]
         {\it Contours for the ``measured values''
             of the total 
             cross section (solid line), ${\cal P}^2/{\cal Q}$ (dashed line), 
             and ${\cal P}^2/{\cal Y}$ (dot-dashed line) 
             in the $[\cos2\phi_L,\cos2\phi_R]$ plane 
             for the set \mbox{\boldmath $RR1$}
             $[\tan\beta=3,~m_0=100~{\rm GeV},
             ~M_{1/2}=200~{\rm GeV}]$ at the $e^+e^-$ c.m. energy of 400 GeV.} 
\label{fig5}
\end{center}
\end{figure}

\vskip -3cm
\begin{figure}
 \begin{center}
\hbox to\textwidth{\hss\epsfig{file=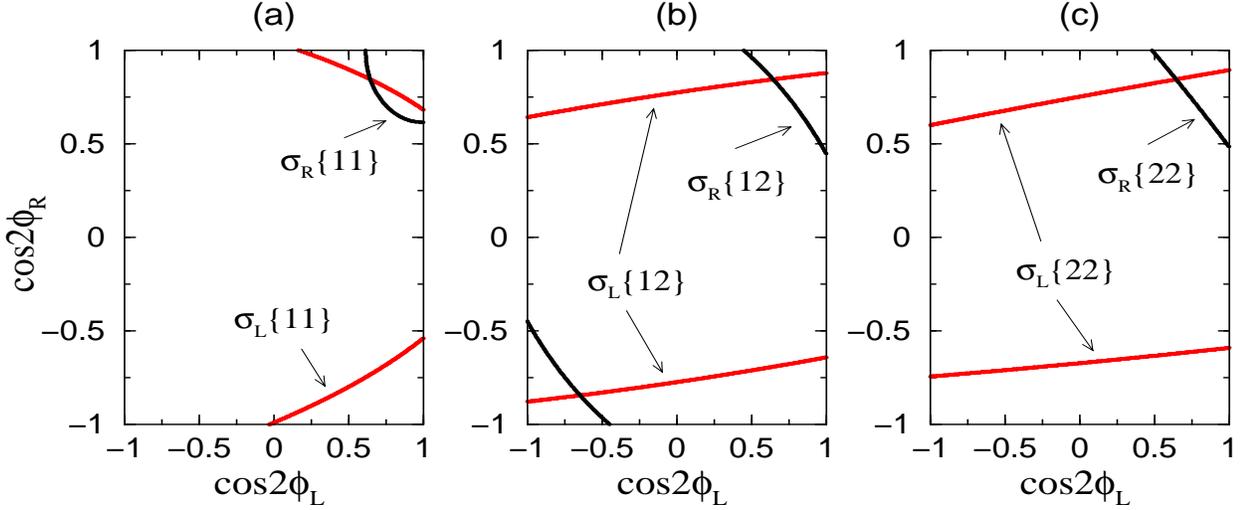,width=16.5cm,
                               height=7cm}\hss}
 \end{center}
\vskip -1cm
\caption[{\bf Figure 6:}]
         {\it Contours of the cross sections 
         (a) $\{\sigma_R\{11\},\sigma_L\{11\}\}$,
         (b) $\{\sigma_R\{12\},\sigma_L\{12\}\}$, and 
         (c) $\{\sigma_R\{22\},\sigma_L\{22\}\}$ 
         in the $[\cos2\phi_L,\cos2\phi_R]$ plane 
         for the set 
         \mbox{\boldmath $RR1$} $[\tan\beta=3,~m_0=100~{\rm GeV},
         ~M_{1/2}=200~{\rm GeV}]$ at the c.m. energy of 800 GeV.}
\label{fig6}
\end{figure}

\vfil\eject

\end{document}